\documentclass[prl,showpacs,twocolumn,amsmath,amssymb,floatfix]{revtex4-1}

\usepackage{graphicx}
\usepackage{bm}
\usepackage{color}
\usepackage[applemac]{inputenc}
 
\newcommand{\sect}[1]{\emph{#1.---}\ignorespaces}

\newcommand{\highlight}[1]{#1}
  
\begin{document} 
   
\title{Mapping the topological phase diagram of multiband semiconductors with supercurrents}

\author{Pablo San-Jose$^1$, Elsa Prada$^2$, Ram\'on Aguado$^1$}
\affiliation{$^1$Instituto de Ciencia de Materiales de Madrid, Consejo Superior de Investigaciones Científicas (ICMM-CSIC), Sor Juana Inés de la Cruz 3, 28049 Madrid, Spain\\$^2$Departamento de Física de la Materia Condensada, Instituto de Ciencia de Materiales Nicolás Cabrera and Condensed Matter Physics Center (IFIMAC), Universidad Autónoma de Madrid, Cantoblanco, 28049 Madrid, Spain}

\date{\today} 

\begin{abstract}
We show that Josephson junctions made of multiband semiconductors with strong spin-orbit coupling carry a critical supercurrent $I_c$ that contains information about the non-trivial topology of the system.  In particular, we find that the emergence and annihilation of Majorana bound states in the junction is reflected in strong even-odd effects in $I_c$ at small junction transparency. 
%This effect is particularly clear when the junction width is smaller than the spin-orbit length, $W/l_\mathrm{so}<1$.
%, such that the system goes from symmetry class D towards class BDI. 
This effect allows for a  mapping between $I_c$ and the  topological phase diagram of the junction, thus providing a dc measurement of its topology.
%This effect allows for a direct dc measurement of the topological phase diagram of the junction through the critical current.
\end{abstract}

\maketitle
%Zero-energy bound states emerging in topological superconductors have recently attracted a great deal of attention owing to their Majorana character. Much of the excitement comes from the possibility of obtaining topological superconductivity via the proximity effect by combining conventional s-wave superconductors with topological insulators or semiconductors with strong spin-orbit (SO) coupling. This can be realized using realistic materials, a discovery that has spurred a race towards the creation and detection of Majorana fermions in condensed matter. 
%Semiconducting nanowires, in particular,  have various experimental advantages as compared with early proposals. The most important is probably the possibility of tuning the topological transition by external knobs. %, such as gate voltages and external magnetic fields. 
%Theory predicts that above a critical external magnetic field $B$, a single-mode wire undergoes a topological transition into a phase hosting zero energy Majorana bound states (MBSs) at the ends of the wire \cite{Lutchyn:PRL10,Oreg:PRL10}. Such transition occurs when the Zeeman energy $V_Z\equiv\frac{1}{2}g\mu_B B$, with $g$ the effective g-factor and $\mu_B$ the Bohr magneton, is larger than $V_Z^c\equiv\sqrt{\mu^2+\Delta^2}$, defined in terms of the Fermi energy $\mu$ and the induced s-wave pairing $\Delta$. %Moreover, nanowires have a significantly enhanced minigap that protects the MBSs from unwanted excitations.  

\highlight{The broad appeal of topological materials lies in their gapped bulk surrounded by exotic edge states \cite{Qi:RMP11,Hasan:RMP10}. 
In topological superconductors, these edge states are Majorana fermions. A particularly interesting case arises when these Majorana fermions are zero-energy bound states \cite{Kitaev:PU01}. Apart from their fundamental interest, these Majorana bound states (MBSs) are fascinating owing to their non-Abelian braiding properties \cite{Read:PRB00} that have been proposed for fault-tolerant, topological, quantum computation. 
New possibilities to engineer topological superconductors, that are more practical than early proposals based on exotic superconductors \cite{Read:PRB00}, have spurred great excitement. These are based on the proximity effect between a conventional s-wave superconductor and a topological insulator \cite{Fu:PRL08}, or a semiconductor wire  with strong spin-orbit (SO) coupling \cite{Sato:PRL09,Sau:PRL10,Alicea:PRB10,Lutchyn:PRL10,Oreg:PRL10}. For the latter %possibility requires the application of a magnetic field $B$ parallel to the wire, that induces a Zeeman splitting $V_Z\equiv\frac{1}{2}g\mu_B B$, with $g$ the effective g-factor and $\mu_B$ the Bohr magneton.
case it has been shown \cite{Lutchyn:PRL10,Oreg:PRL10} that if an external Zeeman field $V_Z$, orthogonal to the SO axis, exceeds a critical value $V_Z^c\equiv\sqrt{\mu^2+\Delta^2}$, where $\mu$ is the Fermi energy and  $\Delta$ the induced s-wave pairing, zero energy MBSs emerge at the wire's ends signaling a topologically non-trivial phase.}

Recent experiments \cite{Mourik:S12}  in InSb nanowires have reported an emergent  zero-bias anomaly (ZBA) in differential conductance $dI/dV$ as Zeeman increases, which is consistent with the existence of  \highlight{zero energy} MBSs \cite{Sengupta:PRB01,Bolech:PRL07,Law:PRL09,Prada:PRB12}. %The main result of these experiments is an emergent zero-bias anomaly (ZBA) in differ$dI/dV$, as $B$ increases, which can be understood as originating from tunneling to the MBS \cite{Sengupta:PRB01,Bolech:PRL07,Law:PRL09}. 
Subsequent experiments \cite{Deng:NL12,Das:NP12} show similar phenomenology, at least superficially. \highlight{The question still stands, however, of whether these constitute true Majorana detections, since other possible explanations}, such as disorder \cite{Pientka:PRL12,Bagrets:PRL12,Liu:PRL12}, smooth confinement \cite{Stanescu:PRB13,Rainis:PRB13,Kells:PRB12,Prada:PRB12}, Kondo physics \cite{Lee:PRL12, Finck:PRL13} (or the related 0.7 structure \cite{Churchill:PRB13}) cannot be completely discarded. Recently, spin-split Andreev levels have been shown to lead to ZBAs similar to those expected for MBSs \cite{Lee:NN14}. It has thus becomes urgent to study alternative ``smoking gun" experiments. 

One possibility is to consider topological Josephson junctions where the fusing of MBSs across the junction results in an odd number of Andreev level crossings at the Fermi energy when the superconducting phase advances by $2\pi$. This is a manifestation of a non-trivial ground state topology, and translates into an anomalous $4\pi$-periodic Josephson effect owing to fermionic parity conservation \cite{Kitaev:PU01,Fu:PRB09,Kwon:EPJB03}. 
However, %recent studies have demonstrated that  
realistic effects that either spoil parity conservation (quasiparticle poisoning) or make the number of crossings even (finite length), translate into a $2\pi$-periodic steady state current  \cite{Badiane:PRL11, San-Jose:PRL12a,Pikulin:PRB12}. Other dc transport observables, however, such as the dissipative multiple Andreev reflection current, or the critical supercurrent $I_c$, retain strong signatures of the drastic changes that the Andreev bound states (ABSs) undergo across the topological transition in the single-band case \cite{San-Jose:NJP13}. 
It is thus natural to ask whether $I_c$ contains information about topology in the general multiband case.

%Nevertheless, the drastic changes that the Andreev levels %spectrum 
%undergo across the topological transition manifest in dc quantities such as the multiple Andreev reflection currents and the critical supercurrent $I_c$ in the single-band case \cite{San-Jose:NJP13}. 
%Thus, it is natural to ask whether $I_c$ contain information about topology in the general multiband case.% of the wire. 

\begin{figure}
   \centering
   \includegraphics[width=\columnwidth]{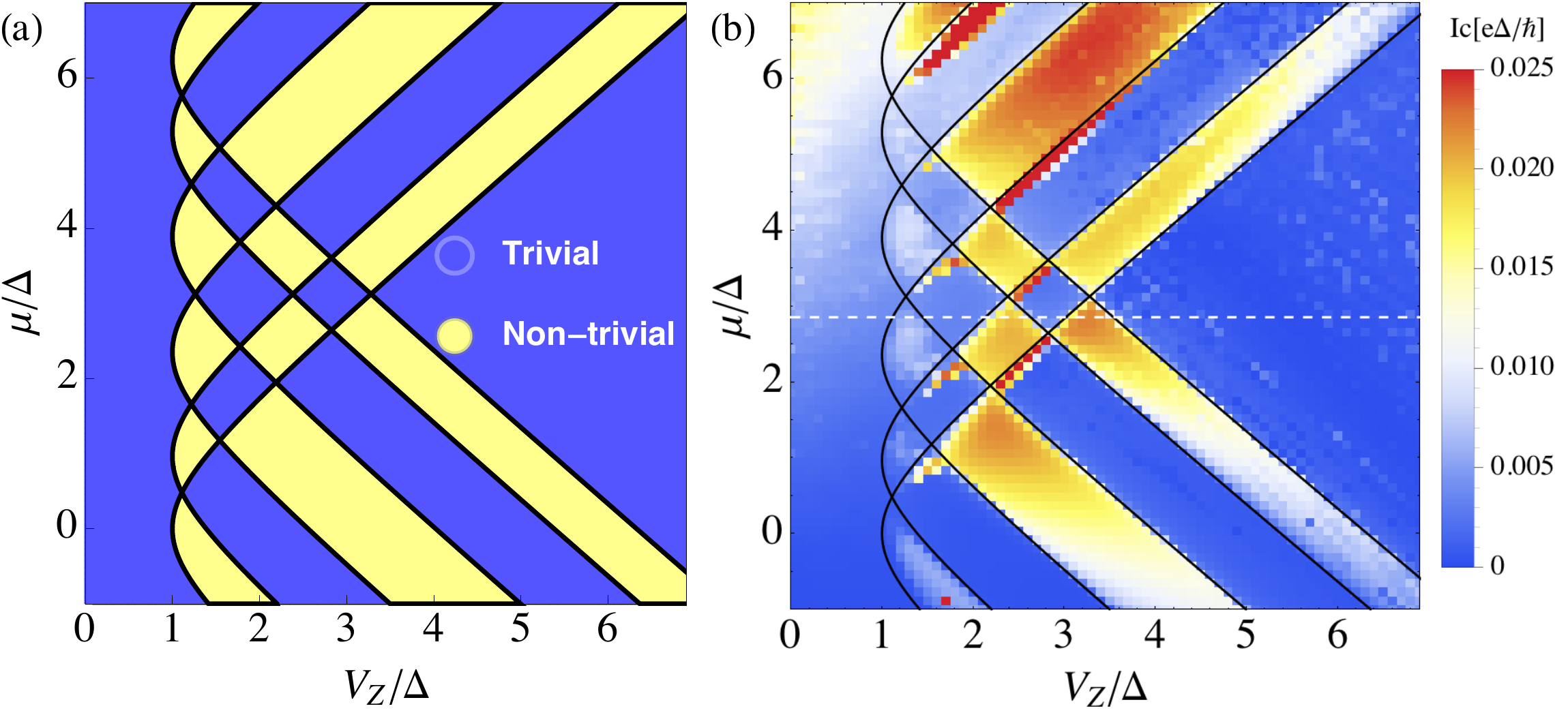} % requires the graphicx package
   \caption{(Color online) (a) Topological phase diagram of a multimode semiconductor wire of width $W=2 l_\mathrm{SO}$, as a function of Zeeman splitting $V_Z$ and Fermi energy $\mu$. (b) Numerical computation of the critical supercurrent across a Josephson junction of small transparency made in such a wire.
   %\highlight{Similar maps are found whenever the Majorana splitting due to subband mixing $\epsilon_0^{(1)}$ exceeds the Majorana hybridisation across the junction  $\delta E_\mathrm{top}^{(1)}$.}
   }
   %(b) Local density of states at the end of the wire for $\mu=0$ (white is zero, and black is maximum). Zero energy states in (b) appear at non-trivial regions of the phase diagram. The red dashed lines mark the Zeeman fields used in Fig. \ref{fig:ABS}.} 
   %(b) Spectrum of the same wire but with finite-length $L_S=10W$. The first Majorana mode appearing at $V_Z=\Delta$ has an associated localization length \cite{Klinovaja:PRB12} $\xi_M\approx\frac{\alpha_{SO}}{\Delta}\approx 0.01L_S$, which explains why this mode does not show oscillations.}
   \label{fig:topoIc}
\end{figure}

\highlight{Here we demonstrate that the $I_c$ of multimode junctions in the tunneling regime indeed provides a measure of topological order.}
In such regime, $I_c(\mu,V_Z)$ maps the full topological phase diagram of the system, see Fig. \ref{fig:topoIc}. It exhibits strong even-odd effects, analogous to those predicted to occur in the spectrum of quasi-one dimensional (quasi-1D) semiconductors containing near-zero MBSs owing to interband Rashba coupling  \cite{Lutchyn:PRL11,Potter:PRB11,Stanescu:PRB11,Gibertini:PRB12}. We also quantify the resilience of the $I_c$ even-odd effect by considering a range of junction transparencies, finite wire length and temperature.
 
\sect{Spectral even-odd effect} The Bogoliubov-de Gennes Hamiltonian of a two-dimensional (2D) Rashba semiconductor reads
\begin{eqnarray}
H&=&\left(\frac{p^2}{2m^*}-\mu\right)\tau_z+\frac{1}{\hbar}\left(\alpha_x\sigma_y p_x\tau_z-\alpha_y\sigma_x p_y\right)\nonumber\\
&+&\Delta\sigma_y\tau_y+V_Z\sigma_x\tau_z,
\label{H}
\end{eqnarray}
where $p^2=p_x^2+p_y^2$, $m^*$ is the effective mass, $\sigma_i$ are the spin Pauli matrices and $\tau_i$ are Pauli matrices in the electron-hole sector. The last two terms describe an induced superconducting pairing of strength $\Delta$ and the Zeeman energy produced by an external magnetic field. For pure Rashba SO, the two couplings $\alpha_{x,y}$ are equal, $\alpha_x=\alpha_y=\alpha_\mathrm{SO}$. For the sake of discussion, we also consider  the possibility that they be different.  
It has been shown \cite{Fujimoto:PRB08, Alicea:PRB10} that the s-wave pairing in Eq. (\ref{H}) induces both an effective  $p_x\pm ip_y$ intraband pairing and an interband s-wave pairing when reexpressed in terms of the $\pm$ eigenbasis of the helical Rashba + Zeeman normal problem. In this new language, a clear physical picture emerges: if $\mu$ and $V_Z$ are such that both $p_x + ip_y$ and $p_x- ip_y$ pairings occur,  the system is topologically trivial. 
In contrast, if $V_Z$ exceeds the critical $V^{c}_Z$, only one of the two p-wave pairings $\Delta_p$ develops. The system then becomes topologically non-trivial. 

In a quasi-1D geometry (with discrete confinement subbands), this picture is replicated for each subband independently (each has its own critical $V^{(n)}_Z=\sqrt{\Delta^2+\mu_n^2}$, with $\mu_n$ the Fermi energy measured from the bottom of the subband). If subbands are not coupled, each will contribute, above its critical $V^{(n)}_Z$, a single zero-energy MBS at each edge. If the subbands are coupled, \highlight{e.g. through the transverse SO coupling $\alpha_y$}, the $N$ MBSs per edge hybridise in pairs  and form $\lfloor N /2 \rfloor$ full fermions at non-zero energies $\epsilon_0^{(n)}$ (with $n=1,\dots \lfloor N /2 \rfloor$) localised at each edge. (We denote the corresponding hole levels by $\epsilon_0^{(-n)}=-\epsilon_0^{(n)}$.) The remaining  $N~\mathrm{mod}~2$ end states stay as Majorana zero modes, at $\epsilon_0^{(0)}=0$ \cite{Lutchyn:PRL11,Potter:PRB11,Stanescu:PRB11}. More formally, the subband coupling destroys the chiral symmetry ($\tau_xH\tau_x=-H$) of the Hamiltonian, reducing its topological invariant from $\mathbb{Z}$ (BDI class in 1D) to $\mathbb{Z}_2$ (D class in 1D) \cite{Schnyder:PRB08,Tewari:PRL12,Altland:PRB97}. This invariant has the meaning of the number of zero energy bound states at the ends of the quasi-1D semiconductor (arbitrary $N$, or $N~\mathrm{mod}~2$). This spectral even-odd effect is thus a manifestation of changes in the topological order in the multiband system.

\begin{figure*}
   \centering
   \includegraphics[width=\textwidth]{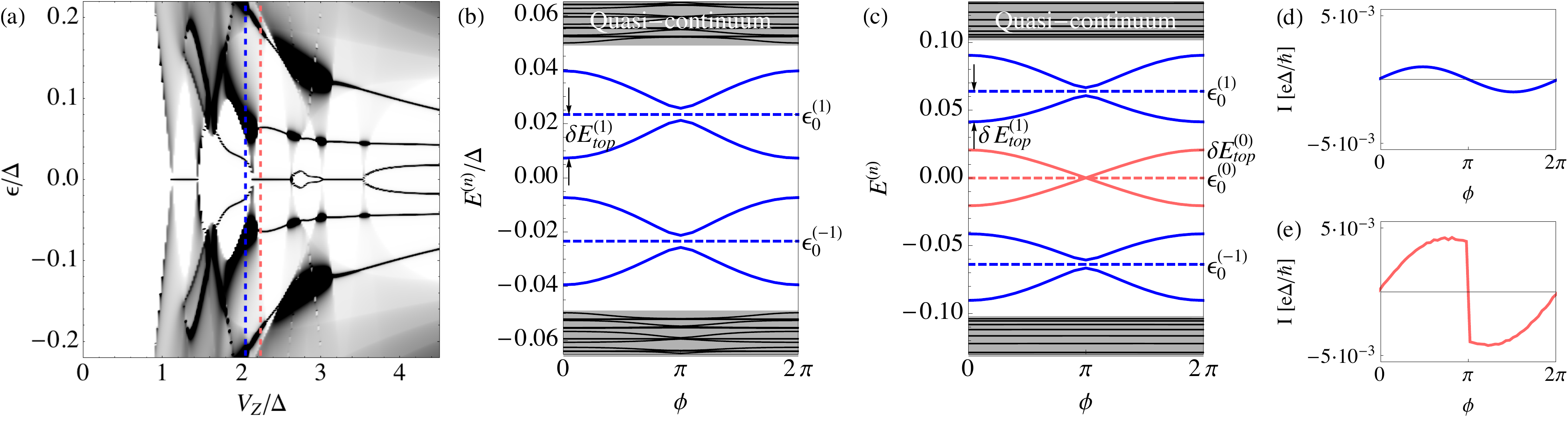} % requires the graphicx package
   \caption{(Color online) (a) Local density of states at the end of a multimode wire for $\mu=2.8\Delta$ (dashed line in Fig. \ref{fig:topoIc}(b)) and width $W=2l_\mathrm{SO}$. White is zero, and black is maximum. The red (light) and blue (dark) dashed lines mark the Zeeman fields $V_Z=2.05\Delta$ and $V_Z=2.25\Delta$ used in panels (b) and (c) respectively. 
(b,c) Andreev energy levels of a multimode Josephson junction as a function of phase difference $\phi$ in a topologically trivial $N=2$ even phase (b), and a non-trivial $N=3$ odd phase (c). The sub-gap levels $\epsilon^{(n)}_0$ at the inner ends of the two wire segments  hybridise across the junction into $\phi$-dependent ABSs of topological origin (red and blue, solid) of amplitude $\delta E_\mathrm{top}^{(n)}$, while those localised at the outer ends remain $\phi$ independent for large wire length $L_S$ (dashed). 
Josephson supercurrents $I(\phi)$ associated to (b) and (c) are shown in (d) and (e) respectively for $T_N=0.001$, including 200 levels into the quasi-continuum.}
   \label{fig:ABS}  
\end{figure*}

To quantify the spectral even-odd effect we analyze the local density of states (LDOS) at the end of a single multimode wire described by Eq. (\ref{H}). We consider a finite value of $W/l_\mathrm{SO}=2$ \footnote{Here $\alpha_x=\alpha_y$, so that $l^x_\mathrm{SO}=l^y_\mathrm{SO}\equiv l_\mathrm{SO}$}, but take the length of the wire $L_S$ to infinity, so as to avoid complications arising from the overlap of Majorana states at opposite edges. The result, for fixed Fermi energy $\mu=2.8\Delta$ [see dotted line in Fig. \ref{fig:topoIc}(b)], as a function of the Zeeman field $V_Z$, is shown in Fig. \ref{fig:ABS}(a). It has been computed within a tight-binding discretisation of Eq. (\ref{H}) using a standard recursive Green's function approach \cite{Datta:97}. At critical fields $V_Z^{(n)}$ each subband becomes topologically non-trivial. These points are visible by a closing of the superconducting gap (light gray), with an accompanying emergence of zero modes for odd transitions (non trivial $\mathbb{Z}_2$ topological order of the wire). The remaining Majoranas are split to higher energies $\epsilon_0^{(n)}$. 
These appear as sharp resonances in Fig. \ref{fig:ABS}(a). 
The odd phases have a zero energy Majorana $\epsilon_0^{(0)}=0$, whilst the $n=0$ is missing in the even phases. 
This phenomenology is replicated throughout the $V_Z-\mu$ plane, resulting in the topological phase diagram shown in Fig. \ref{fig:topoIc}(a) \cite{Lutchyn:PRL11,Potter:PRB11,Stanescu:PRB11}.

\sect{Andreev levels in multiband junctions}
Next we consider a Josephson junction  between two segments in a multimode wire, with a superconducting phase difference $\phi$. Each segment is assumed long (large but finite $L_S$), but the junction is assumed short (normal length $L_N\to 0$), and of a given normal transparency per mode $T_N$. 
If the two segments are decoupled ($T_N=0$), the previous discussion applies to each of them. Sub-gap states of energy $\epsilon^{(n)}_0$ form, localised at both the ``inner" (at the juction) and ``outer" ends of each segment. For finite $T_N$, \highlight{eigenstates in each segment hybridize across the junction, and their energies $E^{(n)}(\phi)$ become $\phi$-dependent. These states are of three types. First, there are states above the gap that are delocalized across the wire, and form a continuum, or a dense quasicontinuum of levels $E_\mathrm{qc}^{(n)}(\phi)$ if the wire is of finite length. Secondly, a discrete set of conventional Andreev sub-gap levels $E_\mathrm{cnv}^{(n)}(\phi)$ form at the junction, two (in general Zeeman-split) for each of the subbands that have not transitioned into their topological phase. Lastly, there are pairs of ABSs of topological origin $E_\mathrm{top}^{(n,\pm)}(\phi)$, resulting from the bonding (-) and antibonding (+) hybridisation of each subgap $\epsilon^{(n)}_0$ on each side of the junction. (For $n=0$, $\pm$ is also the ground state parity). Their $\phi$-dependence at small $T_N$ takes a simple form, $E^{(n, \pm)}_\mathrm{top}(\phi)\approx \epsilon^{(n)}_0\pm \delta E^{(n)}_\mathrm{top} \cos(\phi/2)$ \cite{Kwon:EPJB03,Fu:PRB09}. Note that for large enough $L_S$, the $\epsilon^{(n)}_0$ subgap states localised at the \emph{outer} ends of the wire remain unchanged when increasing the junction transparency. 
We compute all these $E^{(n)}(\phi)$ levels numerically by exact diagonalization of a discretised Eq. (\ref{H}), and show them in Fig. \ref{fig:ABS}(b,c) for a small $T_N$. In the even phase [panel (b)], only $n=\pm 1,\pm 2,\dots$ ABSs of topological origin are present (blue [dark] curves). In the odd phase [panel (c)], an additional $n=0$ ABS appears (red [light] curves) due to the hybridization of the Majorana zero modes. The $\phi=\pi$ crossing is only protected for this $n=0$. The $\phi$-independent levels at energies $\epsilon_0^{(n)}$ (dashed lines) correspond to Majoranas in the outer ends of the segments, while the rest of weakly $\phi$-dependent levels are either the $E_\mathrm{cnv}^{(n)}$ or $E_\mathrm{qc}^{(n)}$, whose variation with $\phi$ vanishes as $T_N$. }

The Majorana hybridisation across the junction was shown, in the case of a single mode,  to be $\delta E^{(0)}_\mathrm{top}=\Delta \sqrt{T_N}$ \cite{Kwon:EPJB03,Fu:PRB09}. In the multimode case, level repulsion makes the $\delta E^{(n)}_\mathrm{top}$ somewhat smaller and weakly $n$-dependent, but it remains true that $\delta E^{(n)}_\mathrm{top}\propto\sqrt{T_N}$ as $T_N\to 0$. \highlight{Hence, in this limit, the $\phi$-amplitude of the topological  $E^{(n, \pm)}_\mathrm{top}(\phi)$ vanishes much more slowly than for conventional levels $E^{(n)}_\mathrm{cnv, qc}(\phi)$}. On the other hand, \highlight{their average $\epsilon^{(n)}_0$ depends on the degree of subband mixing, that may be controlled via the ratio $W/l_\mathrm{SO}$ if the dominant source of mixing is $\alpha_y$. The ratio $\delta E^{(1)}_\mathrm{top}/\epsilon^{(1)}_0$, which will become relevant below,} can therefore be tuned through the junction width and its  transparency. We now show how, depending on this ratio, an analog of the spectral even-odd effect may develop in the junction's $I_c$.

\sect{Even-odd effect in the critical supercurrent}
We consider the $I_c$ of the junction, and how it is affected by the system's topology.
A Josepshon junction supports a supercurrent that is in general given by \cite{Beenakker:92}
\[
I(\phi)=-\frac{1}{2}\frac{e}{\hbar}\sum_{n}\tanh\left[\frac{E^{(n)}(\phi)}{2k_BT}\right]\partial_\phi E^{(n)}(\phi), 
\]
where \highlight{the sum over levels $E^{(n)}$ includes both the standard $E^{(n)}_\mathrm{cnv, qc}$ and the $E^{(n,\pm)}_\mathrm{top}$ of topological origin. In the tunnelling limit, the sum is dominated by the latter, while the former, due to their suppressed $\phi$-derivative, contribute with a weak and featureless background, even across topological transitions.} 

Typical Josephson currents $I(\phi)$ in the tunnelling limit for even and odd phases are shown in Fig. \ref{fig:ABS}(d,e). Note that although both are $2\pi$ periodic \cite{Badiane:PRL11, San-Jose:PRL12a,Pikulin:PRB12}, \highlight{phases with non-trivial topology exhibit} a sawtooth $I(\phi)$ profile at small transparencies, panel (e). The critical current $I_c$ is given by the maximum of $|I(\phi)|$. If $\delta E_\mathrm{top}^{(1)}<\epsilon^{(1)}_0$, then $\delta E_\mathrm{top}^{(n)}<\epsilon^{(n)}_0$ for all $n>0$. In this regime, the zero-temperature contribution to $I(\phi)$, and therefore to $I_c$, of $E^{(n,+)}_\mathrm{top}$ and $E^{(n,-)}_\mathrm{top}$ will approximately cancel (to order $\sqrt{T_N}$) for any $n>0$ and $\phi$ (they have opposite $\phi$-derivative, and remain of constant sign for all $\phi$). However, the $n=0$ Majorana ABS, if present, does not cancel due to the sign change of the $\tanh$ function at zero energy. This extra contribution produces a $\sim \sqrt{T_N}$ increase in $I_c$ in odd phases,
\begin{equation}
I_c\approx I_c^\mathrm{bg}+\left\{\begin{array}{ll}
0&\textrm{if $N$ mod $2=0$}\\
\frac{1}{2}\frac{e}{\hbar} \delta E_\mathrm{top}^{(0)}&\textrm{if $N$ mod $2=1$}
\end{array}
\right.\space (\textrm{for } \delta E_\mathrm{top}^{(1)}<\epsilon^{(1)}_0),
\label{Ic}
\end{equation}
\highlight{where the background $I_c^\mathrm{bg}$ from conventional ABSs and quasicontinuum levels $E^{(n)}_\mathrm{cnv, qc}$ is small ($\sim T_N$) and independent of the system's $\mathbb{Z}_2$ topological invariant $N$ mod 2. The $I_c$ even-odd constrast grows like $T_N^{-1/2}$ as $T_N\to 0$.}

\begin{figure}
   \centering
   \includegraphics[width=\columnwidth]{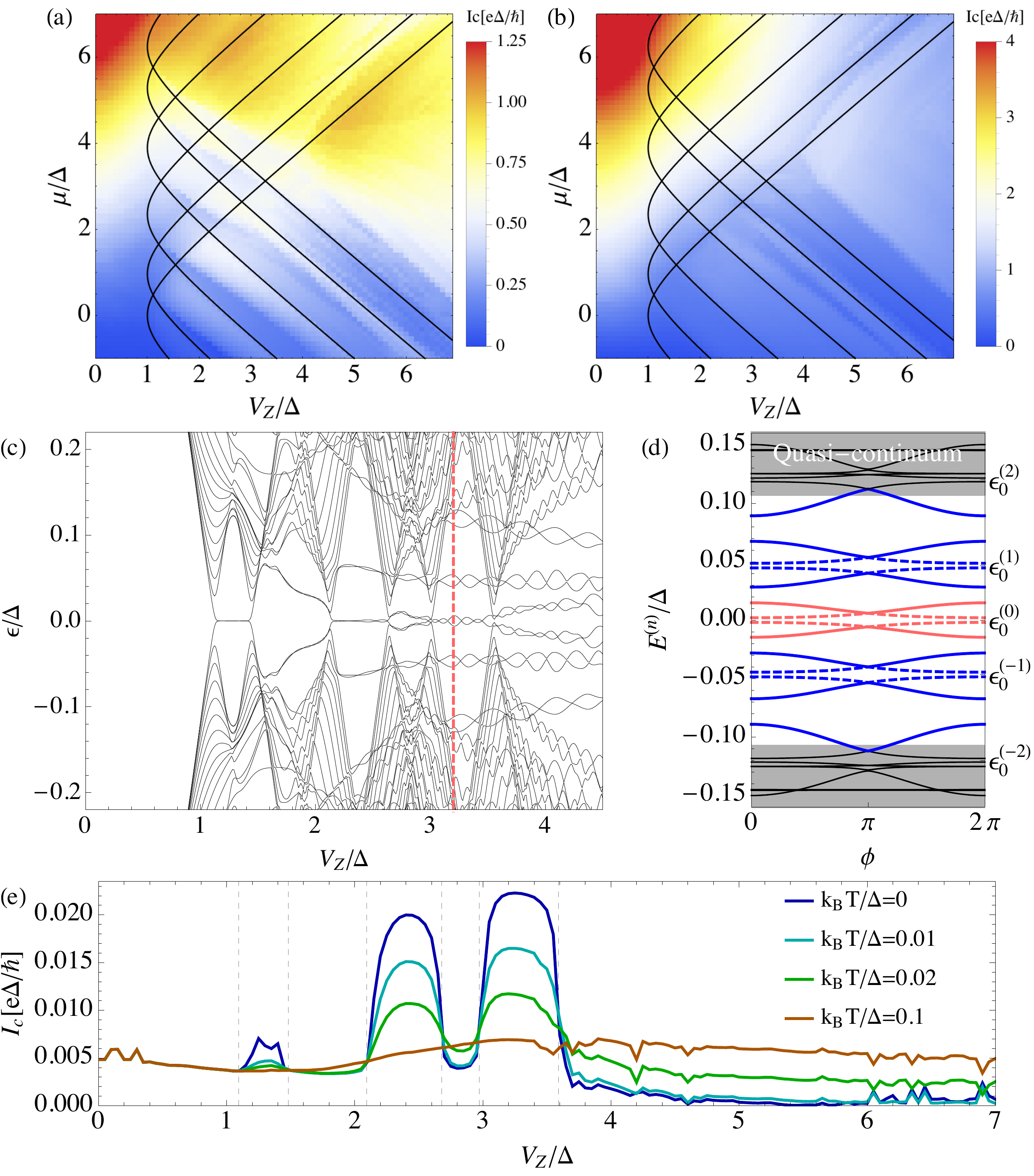} 
   \caption{(Color online) (a,b) Zero temperature $I_c$ like in Fig. \ref{fig:topoIc}(b) for \highlight{$T_N=0.5$, (a),  and $T_N=1$, (b)}. (c) Spectrum of a single wire like that of Fig. \ref{fig:ABS}(a), but with a finite length $L_S=8.7W$. (d) Corresponding ABS spectrum at $V_Z=3.2\Delta$ [dashed line in (c)] at $T_N=0.005$. (e) Cut of the $I_c$ map of Fig. \ref{fig:topoIc}(b) for $\mu=2.8\Delta$ at different temperatures.}
   \label{fig:deviations}
\end{figure}

The result above suggests a strong signature of the spectral even-odd effect should be visible in $I_c$ at small $T_N$. This is confirmed by the numerical simulation shown in Fig. \ref{fig:topoIc}(b), where the zero temperature map $I_c(\mu, V_Z)$ is computed with 200 ABSs (convergence achieved), obtained by exact diagonalization of a six-subband Josephson junction with $W/l_\mathrm{SO}=2$ and $T_N=0.005$ such that $\delta E_\mathrm{top}^{(1)}<\epsilon^{(1)}_0$ \footnote{For the plots of Fig. \ref{fig:topoIc}(b) and \ref{fig:deviations}(a,b) we have considered $\alpha_x\neq \alpha_y$, particularly $\alpha_y/\alpha_x=0.4$. We introduce this artificial asymmetry for computational reasons. With a stronger SO coupling in the $x$-direction, we can perform our numerical calculations with a shorter total length of the system such that the outer Majoranas remain decoupled from the junction, yielding flat ABSs as a function of $\phi$. It should be noted, however, that the even-odd effect we describe is independent of this asymmetry (as long as inner and outer Majoranas do not overlap).}. Critical fields $V_Z^{(n)}$ are marked with black lines. Note that a large contrast in $I_c$ is visible between trivial and non-trivial phases. 
A cut for $\mu=2.8\Delta$ is shown in Fig. \ref{fig:deviations}(e), blue (upper) curve.

A number of factors suppress this correspondence between $I_c$ and topology. The most immediate are a high normal transparency of the junction \highlight{(which decreases the contrast and increases the amplitude $\delta E_\mathrm{top}^{(1)}$) and a smaller width (which decreases $\epsilon_0^{(1)}$)}. As soon as $\delta E_\mathrm{top}^{(n)}>\epsilon^{(n)}_0$, the cancellation of supercurrents from $E^{(n,+)}_\mathrm{top}$ and $E^{(n,-)}_\mathrm{top}$ becomes restricted to a finite range around $\phi=\pi$, where the sign of the $\tanh$ is constant. Outside of this range, the two levels will add instead of cancel, and \highlight{their contribution to} $I_c$ will become finite even for $N$ mod 2=0. \highlight{The resulting loss of topological contrast is illustrated in Fig. \ref{fig:deviations}(a,b), where the same map of Fig. \ref{fig:topoIc}(b) is presented for $T_N=0.5$ and $T_N=1$, respectively.}

Another potentially deleterious factor to the $I_c$ even-odd effect is the finite length $L_S$ of the two superconducting segments of the wire. In a single decoupled wire, the MBSs at each end decay into the bulk within a typical localisation length \cite{Klinovaja:PRB12} of the order of the coherence length of each subband  $\xi_n\approx \hbar v_F^{(n)}/\Delta$ (where $v_F^{(n)}$ is the corresponding Fermi velocity).
When $L_S\lesssim \max_n \xi_n$, the Majorana states of different ends overlap. This modifies the spectral even-odd structure, since zero modes at different ends in an odd phase hybridise into a fermion of finite energy that oscillates with $V_Z$ \cite{Lim:PRB12, Prada:PRB12,Rainis:PRB13,Das-Sarma:PRB12}.  Fig. \ref{fig:deviations}(c) shows the spectrum of a wire like in Fig. \ref{fig:ABS}(a), but with $L_S=8.7\;W$. Note, however, that while a finite $L_S$ leads to zero-mode splittings at high fields $V_Z$, those at smaller fields are typically much more robust, owing to their smaller Fermi velocities (which grows with $V_Z$) and hence smaller localization lengths.
%Although the Majorana zero modes at each edge remain approximately at zero energy in the odd phase below $V_Z^{(4)}$ (for which $\xi_1$ is smaller than $L_S$), those at higher values of $V_Z$ do not, since their localisation length (through the associated Fermi velocities) increases with $V_Z$.
In a Josephson junction geometry, a finite $L_S$ will likewise lead to the hybridisation of outer- and inner-edge zero modes, see Fig. \ref{fig:deviations}(d).
This suppresses their contribution to the supercurrent and hence the even-odd contrast in $I_c$, but typically only for high $V_Z$ like in panel (c). Note, moreover, that even though the spectrum in Fig. \ref{fig:deviations}(d) corresponds to an odd phase, it is actually trivial for finite $L_S$, rigorously speaking \cite{Pikulin:JL11}, since it has an even number of zero energy crossings as a function of $\phi$ \cite{Beenakker:PRL13}.

% Although this junction is nominally nontrivial according to the phase diagram [dashed line in (c), $\mu=2.8\Delta$, $V_Z=3.2\Delta$], the ABS spectrum has an even (zero) number of zero energy crossings as a function of $\phi$, indicating \cite{Beenakker:PRL13} that, rigorously, the topology of the junction is trivial \cite{Pikulin:JL11}.

\highlight{Finally, we consider the effect of finite temperature}. Fig. \ref{fig:deviations}(e) shows a cut of the $I_c$ map of Fig. \ref{fig:topoIc}(b) for $\mu=2.8\Delta$ and four different temperatures. Vertical dashed lines mark the critical fields $V_Z^{(n)}$. Note that $I_c$ exhibits a smooth background offset $I_c^\mathrm{bg}$, with comparatively large plateaus superimposed in odd phases [see the blue (upper) curve for $T=0$]. Increasing temperatures above $\epsilon_0^{(1)}$ quickly washes out \highlight{the even-odd contrast}. The energy $\epsilon_0^{(1)}$ [around $0.07 \Delta$ in this example, see Fig. \ref{fig:ABS}(c)] is sometimes called the mini gap \cite{Stanescu:PRB11}, or the energy scale protecting Majorana physics, and consequently the $I_c$ even-odd effect.

In conclusion, we have shown that the critical supercurrent of a multimode Josephson junction in a  quasi-1D Rashba semiconductor wire with induced superconductivity, \highlight{can} be used to directly measure the $\mathbb{Z}_2$ topological invariant. This invariant gives the number -- zero (trivial) or one (non-trivial) -- of Majorana zero modes at each boundary in the wire. We found that, in topologically non-trivial phases, the $I_c$ \highlight{for small junction transparency is strongly enhanced relative to trivial phases,}
by virtue of the additional supercurrent contributed by Majorana zero modes in the junction. It represents a transport analogue of the spectral even-odd effect in a uniform wire.
Given the extremely clean and precise measurements of $I_c$ possible today \cite{Amado:PRB13,Deng:NL12,Doh:S05,Nishio:N11,Nilsson:NL12,Gunel:JOAP12}, this effect could prove to be a useful route towards the detection of topological transitions, and hence MBSs in these systems, and an alternative to the ZBAs exploited to date.
We have evaluated the impact of a range of factors that may mask the even-odd contrast, and concluded that ballistic junctions currently available should exhibit topological signatures in $I_c$ at temperatures below a few hundreds of mK and small transparency.

We gratefully acknowledge discussions with F. Giazotto, M. Amado and A. Fornieri. This work was supported by the European Research Council and the Spanish Ministry of Economy and Innovation through Grants No. FIS2011-23713 (P.S.-J), FIS2012-33521 (R.A.), No. FIS2010-21883 and the Ram\'on y Cajal Program (E. P). 

\bibliography{biblio}

\end{document}